# Title

On gauge freedom and cooperative self organization in glassy As$_2$Se$_3$

C.B. Nelson
Department of Physics and Astronomy
St. Cloud State University

## Abstract

The gauge principle as applied in field theories for particle physics recognizes redundant degrees of freedom.  Some time ago this author introduced a frustration based locally preferred structure (LPS) model of the glass transition. Here we show that gauge freedom plays an essential role in the self organized emergence of two temperature dependent quantities $\xi_\beta(T), \xi_\theta(T)$, which connect the time dependence of the bond angle fluctuations $\delta\beta_i, \delta\theta_{ij}$ as a given LPS approaches the glass temperature. We show that as a choice of gauge these quantities cooperatively converge as $\xi_\beta(T_g)^{-1} \approx \xi_\theta(T_g)^{-1} \approx 0.6$, which is equivalent to the stretched exponential power observed in structural glasses.  We therefore posit that gauge freedom plays a deeper role in the self-organization of this non-equilibrium system.

## I. Introduction

In 2014 J.S Langer published a review article [1] stating that there is no generally accepted fundamental understanding of glassy states in matter or the processes by which they are formed. Today, as far as we know, this situation persists. Two glass models of current interest are the shear-transformation-zone (STZ) model [2,3,4], and the random first order phase transition (RFOT) model [5,6,7]. Both of these models have seen wide application at macroscopic size scales. The STZ model posits that local events, including noise and shear forces control the dynamics. This model does not include local frustration, which has been recognized as playing a crucial role in glass dynamics [1].

Some time ago this author introduced a local preferred structure model (LPS) for $As_2Se_3$ glass. This model is based on a microscopic tight binding (TB) statistical mechanical approach [8]. We posit that the glass transition can be explained locally by fluctuations in two bond angles $\delta\theta_{ij}, \delta\beta_i$ and in their associated bond lengths, $\delta d_i$ in that are "frozen" into each LPS in the glass. The local dynamics incorporate frustration as a gauge field with local U1 symmetry [9]. As mentioned earlier [10], the magnitudes of the frustration fields are too large, so we also introduce bath modes $q_{\beta i}$ as a random directional field which competes with frustration. These modes in part determine the magnitude of the energy extinction times which are defined later in the introduction. It is well known that pattern formation occurs when non-equilibrium systems freeze [11]. We posit that the NQR distribution is such a formation for this glass. As far we know, neither the STZ or RFOT model have been used to predict the NQR distribution. In addition, the model predicts the peak in the specific heat, glass temperature $T_g$, and the Kauzmann temperature $T_K$.

Now consider the stretched exponential time dependence of the relaxation of this glass, as evidenced in the equation, $\phi(t) = \exp\left[(-t/\tau)^\lambda\right]$. Here t is a laboratory time, $\tau$ is a relaxation time that material dependent, and $\lambda$ is a scalar quantity also material dependent. Several different models have been used to derive $\lambda$ in disordered systems. Among these are the direct transfer model [12], the hierarchically constrained dynamics model [13], and the direct diffusion model [14]. All of these share the generation of a scale invariant distribution of relaxation times [15]. In terms of structural glass, Philips [16,17] and others [18, 19] developed a model used to calculate $\lambda$ based on the diffusion of quasi-particles between randomly distributed traps. These determined a value $\lambda \approx 0.6$, independent of the glass temperature.

It has long been recognized [20] that cooperative phenomenon place a central role in non-equilibrium systems. In the 1980's Per Bak posited that large non-equilibrium open systems subject to random external fluctuations exhibit so called self organized criticality [21,22]. Per Bak asserted that scaling laws exist between certain 1 dimensional quantities at a local level and distributions of these at a non-local level. This was applied to a variety of phenomenon, including earthquakes and the length of the coast of Norway. However, at the end of his text Per Bak asserted that the principle of self organized criticality should apply to other non-equilibrium phenomenon. We posit that if self organized criticality is as deep a "universal phenomenon" as Per Bak suggested it can occur in quantities such as fluctuations in bond angle and associated relaxation times. In an earlier paper we showed that the NQR distribution of the structural glass could be predicted if we assumed that on the order of $10^{15}$ different LPS's exist in a given sample of the glass. We showed these structures differ from the crystal in local bond fluctuations (see Figure 1) $\delta\theta_{ij}, \delta\beta_i$, and local bond lengths as mentioned above. In this model we assert that self organized criticality occurs in the glass with respect to a dimensionless product of times characteristic of a given LPS freezing in. We posit that a fundamental structure exists which is a network of 4 coupled LPS's as shown in Figure 8 in the next section. This is the smallest self similar structure which can be formed from LPS's. We relate this network to a single LPS using a self organization approach to calculate two power law quantities $\xi_{\beta,\theta}(T)$. We then show that $\xi_{\beta,\theta}(T)$ converge independently to a quantity which is proportional to $\lambda^{-1}$, the stretched exponential derived for structural glasses. Development of our approach is begun by considering the figure below:

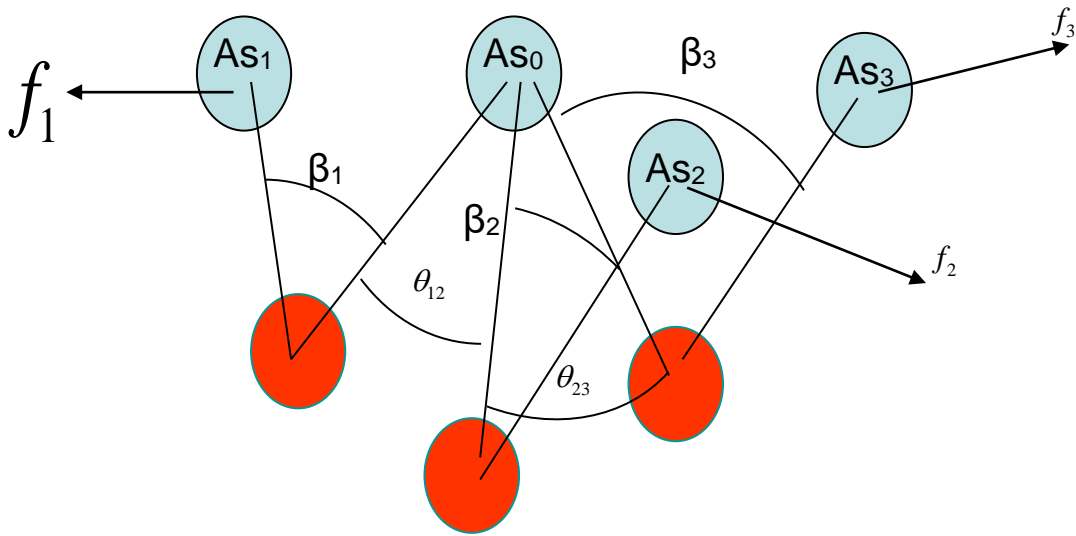

**Figure 1. Local preferred Structure (LPS) of $As_2Se_3$ glass used in this model. The frustration forces re shown along with the hybrid angles $\beta_i$, and two of the bonding angles $\theta_{ij}$.**

In the model [9] we assumed that the bond angles shown above have certain average values which predicts an NQR frequency at the center of the distribution for the glass, which we assume represents a quantum super position of LPS's above the glass temperature. Once the system begins to creep toward $T_g$ by dumping energy to the bath [10] the local frustration forces emerge first, inducing the fluctuations $O_{\beta i} = \sqrt{\delta \beta_i}$. We posited that the field $O_{\beta i}$ has the positional and temporal dependence $O_{\beta i} = O_{0\beta i} e^{-i\alpha_{\beta i}(S_i)} e^{-\gamma_{\beta i} t}$ [10]. The local gauge transformation $\alpha_{\beta i}(S_i)$ is determined by equation (36) in [10]. The fluctuations $\delta \theta_{ij}$ are assumed to emerge slightly later and have the dependence $\delta \theta_{ij} = \delta \theta_{0ij} e^{-\gamma_{\theta ij} t}$ [10]. Both of the fluctuation amplitudes $\delta \theta_{0ij}, O_{0\beta i}$ are determined by fitting the calculated NQR distribution to the data [9], thus collapsing a quantum superposition of LPS's. In this fashion it is seen that the fluctuation $O_{\beta i}$ can be interpreted as a standing wave with an exponentially decreasing lifetime, whose final amplitude $O_{0\beta i}$ gets "frozen in". Since the system is assumed initially to be in a quantum superposition, the inverse relaxation times $\gamma_{\beta i}(T), \gamma_{\theta ij}(T)$ must be functions of temperature [23]. The most straight forward way to accomplish this is write these in terms of probabilities in the thermodynamic sense. Once each bond "freezes in" and the temperature of the system is below $T_g$, we assume each bond is in "local equilibrium", and free energies $\delta F_{\beta i}, \delta F_{\theta ij}$ for each fluctuation are used [9] to write partition functions. These are in turn used to calculate the probabilities. The frustration forces form before each bond fluctuation $\delta \beta$ freezes in [10]. Once this begins to occur it makes no sense to assume all bonds in a particular LPS freeze out simultaneously. We posit that an aspect of co-operation is that the bond fluctuations freeze out sequentially as functions of the ground state energies, $\delta E_{0\beta i}, \delta E_{0\theta_{ij}}$ [10]. A consequence of this is that as a particular LPS freezes the probabilities are multiplied, thus incorporating a type of quantum interference as co-operative behavior [24].

In ref [9] it was shown that $\delta \theta_{ij}, \delta \beta_i$ both couple to particular bath modes $q_{\beta i}, q_{\theta ij}$ through the interaction Hamiltonians given in equations (12, 38). We also assumed that the two bath quanta had a local time dependence, $q_{\beta i} = q_{0\beta i} e^{-\gamma_{q\beta} t}, q_{\theta ij} = q_{0\theta ij} e^{-\gamma_{q\theta} t}$. Here $\gamma_{q\beta}, \gamma_{q\theta ij}$ are relaxation times to be determined. These bath quanta propagate outward through the glass as it nucleates, so a proper description of their behavior may require using non-Markovian dynamics [25, 26, 27], which we don't consider here. We note that since the bath quanta carries energy away from the glass as the fields $\delta\beta_j, \delta\theta_{ij}$ drop into their respective ground states this process is akin to spontaneous emission [28, 29], and we posit $\gamma_{q\beta} \approx \gamma_{\beta i}(T_g), \gamma_{q\theta ij} \approx \gamma_{\theta ij}(T_g)$. Consider that these quanta do not decay away immediately, but rather propagate into the bath with the amplitudes $q_{0\beta i}, q_{0\theta ij}$. Here we posit that $\gamma_{q\beta} \approx i\omega_{q\beta}, \gamma_{q\theta ij} \approx i\omega_{q\theta ij}$, where these angular frequencies are determined by the initial energies the bath quanta are emitted with.

As the glass is "submerged" in a lower temperature bath we assume that the frustration fields [9] are first to form. These can be calculated from the local free energy $F_{\theta ij}$ before the ground states in the fluctuations $\delta\theta_{ij}$ are defined, $\gamma_{\beta i}, \gamma_{\theta ij}$. The time dependence of these fluctuations with respect to the time dependence of the fluctuations $\delta O_{\beta i}$ must be evident in the relative magnitudes of the inverse relaxation times. The time intervals $\gamma_{\beta i}^{-1}, \gamma_{\theta ij}^{-1}$ can be identified as "short" relaxation times [1]. In addition to these, we also define temporal quantities $\Delta t_{\beta i}(T), \Delta t_{\theta ij}(T)$ (energy extinction times) for which the free energies $\delta F_{\beta i}, \delta F_{\theta ij}$ of the fluctuations $\delta\beta_i, \delta\theta_{ij}$ vanish. A given LPS can be represented by the pairs $(\gamma_{\beta i}, \Delta t_{\beta i}), (\gamma_{\theta ij}, \Delta t_{\theta ij})$. Products of these are dimensionless. In the next section we calculate power law relationships between an LPS as defined in terms of products of these pairs, and the larger structure defined in Figure 8. This allows us to calculate $\xi_\beta(T), \xi_\theta(T)$ as functions of temperature. We show that in the limit $\underset{T \to T_g}{Lim} \xi_{\theta,\beta}^{-1}(T) \approx 0.6$, which suggests a connection between temperature dependent cooperative self organization and the stretched exponential behavior of structural glasses.

## II. Model

The calculations we are doing involve structures and their attendant dynamics defined at the atomic/molecular level. One of the key terms that determine the dynamics in both systems is the valence electronic energy $E_0(\beta_i, \theta_{ij}, d_{ij})$ as seen in the fluctuations of the free energies [10] shown below;

$$\delta F_{\beta 0_i} = \left\{ \left( \sum_i^j \frac{\partial E_0(\beta_i)}{\partial \beta_i} - \left( \sqrt{|\vec{f_i}|} + i\sqrt{\hat{f_i} \Box \sum_{j \neq i} \vec{f_j}} \right)^2 \cdot d_i \right) \delta\beta_i - \kappa d_i^2 q_i \delta\beta_i + \frac{\kappa d_i^2}{2} \delta\beta_i^2 \right\}, \quad (1)$$

$$\delta F_{0\theta_{ij}} = \left\{ \frac{\partial E(\theta_{ij})}{\partial \theta_{ij}} \delta\theta_{ij} + \sum_i \frac{\partial \left[ V_{pp\sigma,\pi i} \right]}{\partial \theta_{ij}} \delta\theta_{ij} + \kappa d^2 q \delta\theta_{ij} + \frac{\kappa d^2}{2} \delta\theta_{ij}^2 \right\}. \quad (2)$$

This valence energy consists of tight binding terms which are used to approximate the overlap energies in the bonds of the LPS, and as such contain $\hbar$, so these terms are already quantum mechanical in nature. To obtain the ground state scalar values of the fields $\delta\beta_{0i}, \delta\theta_{0ij}$, we take derivatives,

$$\frac{\partial \delta F_{0\beta_i}}{\partial \delta\beta_i}\bigg|_{\delta\beta_{0i}} = 0, \frac{\partial \delta F_{0\theta_{ij}}}{\partial \delta\theta_{ij}}\bigg|_{\delta\theta_{ii}} = 0. \quad (3.a,b)$$

which, when substituted back into equations (1) & (2) are shown below;

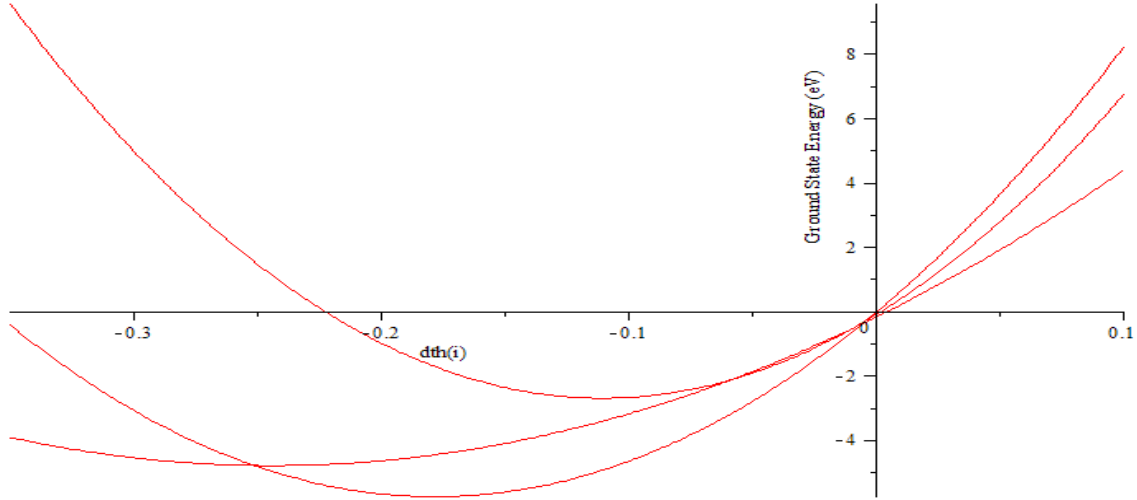

**Figure 2. Ground state energies $E_{0\theta ij}(\delta\theta_{ij})$ of the $\delta\theta_{ij}$ bond fluctuations.**

Notice the $\delta\theta_{ij}$ are negative, which is consistent with the fact that the glass is denser than the liquid. This yields $\delta\theta_{12} = -0.12, \delta\theta_{23} = -0.175, \delta\theta_{13} = -0.25$ (in degrees) which is consistent with our earlier work [9]. As mentioned, it also in part explains why the structural changes are too small to be determined by x-ray or neutron diffraction. The associated ground state energies are
$\delta E_{0\theta 11} = -2.0 eV, \delta E_{0\theta 23} = -3.0 eV, \delta E_{0\theta 13} = -8.0 eV$.

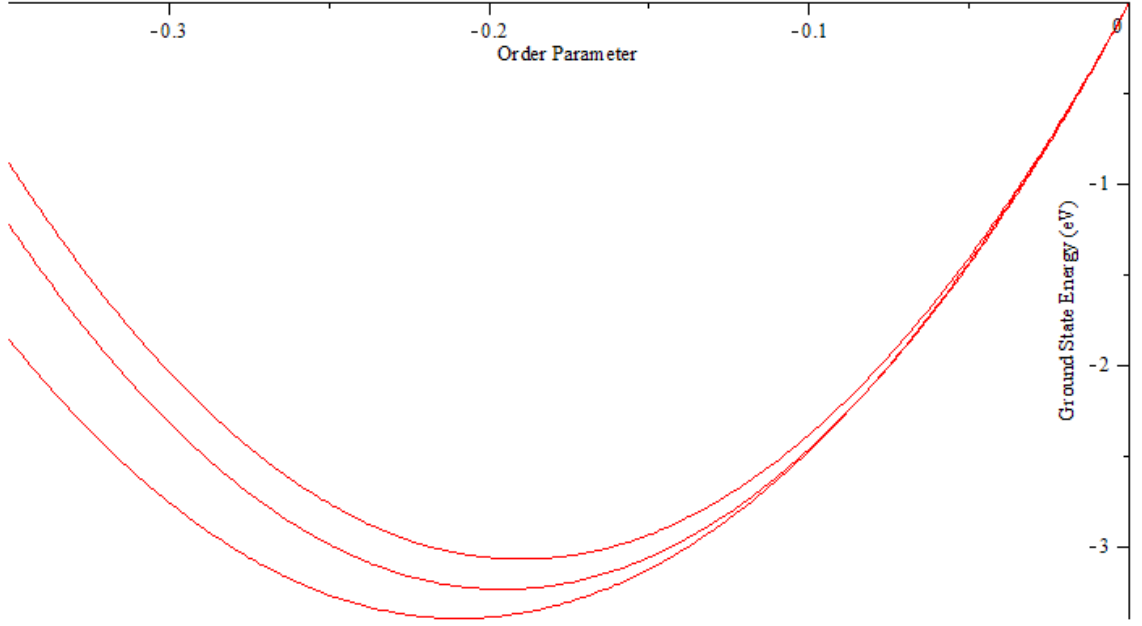

**Figure 3. Ground state energies $E_{0\beta j}(\delta\beta_j)$ of the $\delta\beta_j$ bond fluctuations, which are the order parameters.**

Next we solve equation (3.a) to obtain magnitudes of the order parameters being $\delta\beta_1 = -0.20, \delta\beta_2 = -0.19, \delta\beta = -0.18$. These are also consistent with our earlier work [9], and sine they are negative, they agree with Figure 2 in suggesting the glass condenses as it forms. The associated ground state energies are $\delta E_{0\beta 1} = -3.4 eV, \delta E_{0\beta 2} = -3.0 eV, \delta E_{0\beta 3} = -2.85 eV$, respectively.
In [10] we introduced the inverse relaxation times for the glass fluctuations $\delta\theta_i = \delta\theta_{i0}e^{-\gamma_{i\theta}t}, \delta\beta_j = \delta\beta_{j0}e^{-2\gamma_{j\beta}t}$ as $\gamma_{j\beta}(T), \gamma_{i\theta}(T)$, which are partial functions of temperature. Here i, j are bond indices on a particular (LPS). Also, it is untenable to assume [10] that the relaxation times of these fluctuations are not quantum mechanical. The probabilities of a particular bonding configuration are defined in the usual way as

$$P_{\beta i} = \frac{Z_{\beta i}}{\sum_{j=1}^{3} Z_{\beta j}},$$

$$P_{\theta ij} = \frac{Z_{\theta ij}}{\sum_{i,j=1}^{3} Z_{\theta ij}}.$$
(4.a,b)

Here the Z terms are the partition functions of the frozen bond fluctuations.

In order to determine the forms of the inverse relaxation times we first assume that the temperature dependence of $\gamma_{\beta j}(T), \gamma_{\theta ij}(T)$ manifests through the probabilities, so that $\gamma_{\beta j}(P_{\beta j}(T)), \gamma_{\theta ij}(P_{\theta ij}(T))$. The simplest way to obtain this is to assume

$$\gamma_{\beta j} \approx P_{\beta j}(T), \gamma_{\theta ij} \approx P_{\theta ij}(T).$$
(5.a,b)

This is not unreasonable considering that the formation of these bond fluctuations is non-equilibrium, thus we are in a way assigning a temperature dependence to the inverse times. Secondly we assume the inverse times are proportional to the magnitudes of correlated ground state energies,

$$\gamma_{\beta j} \approx \delta E_{0\beta j}, \gamma_{\theta ij} \approx \delta E_{0\theta ij}.$$
(6.a,b)

Combining these two sets of equations gives us,

$$\gamma_{\beta j} \approx P_{\beta j}(T)\delta E_{0\beta j}, \gamma_{\theta ij} \approx P_{\theta ij}(T)\delta E_{0\theta ij}.$$
(7.a,b)

These have units of energy (eV), so we posit that we may divide each expression by a constant with the units of eV-sec. Since these probabilities as well as the energies are calculated using (TB) theory, we identify this constant as h (Planck's constant). It is our view that a cooperative process requires the probability of the formation of a particular bond be dependent on the just previous formation of an adjacent bond. Considering figure (1) we would have going around clockwise from top,

$$\gamma_{\beta 1} = \frac{P_{\beta 1}\delta E_{0\beta 1}}{h}, \gamma_{\beta 2} = \frac{P_{\beta 1}P_{\beta 2}\delta E_{0\beta 2}}{h}, \gamma_{\beta 3} = \frac{P_{\beta 1}P_{\beta 2}P_{\beta 3}\delta E_{0\beta 3}}{h}.$$
(8.a,b,c)

And similarly with the $\delta\theta_{0ij}$ fields,

$$\gamma_{\theta 12} = \frac{P_{\theta 12}\delta E_{0\theta 12}}{h}, \gamma_{\theta 23} = \frac{P_{\theta 12}P_{\theta 23}\delta E_{0\theta 23}}{h}, \gamma_{\theta 13} = \frac{P_{\theta 12}P_{\theta 23}P_{\theta 13}\delta E_{0\theta 13}}{h}. \quad (9.a,b,c)$$

Consider that the inverse relaxation times determine when the fluctuations freeze out, so the $\gamma_{\beta j}, \gamma_{\theta ij}$ are unique to the material. Below we plot inverse relaxation times given in equations (8).

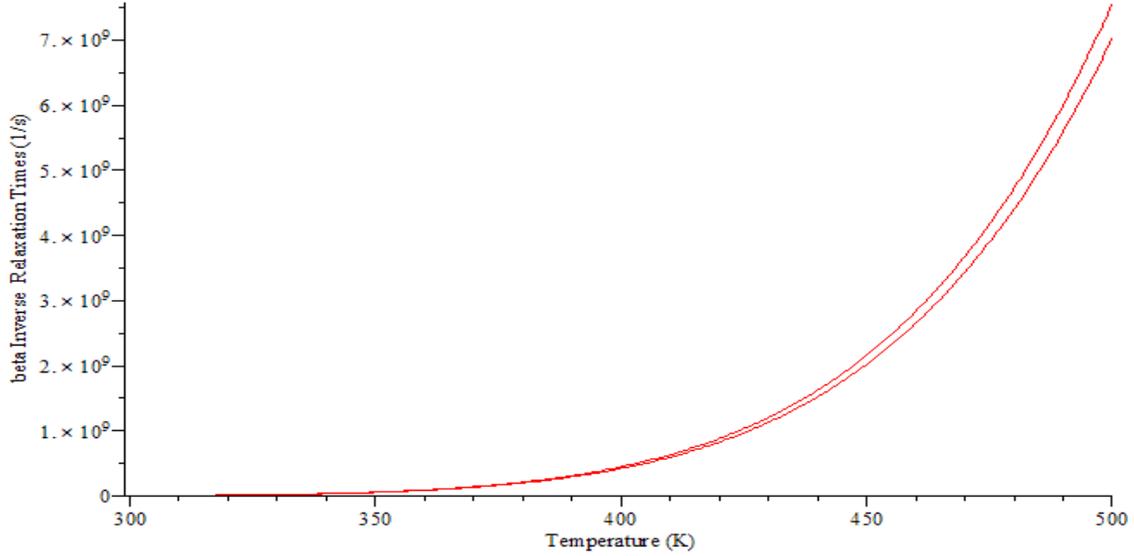

**Figure 4. Inverse relaxation times $\gamma_{\beta 1}, \gamma_{\beta 2}$. Note the strong overlap for these quantities as $T \to 0$.**

Notice that these converge: $\underset{T\to 0}{Lim}\gamma_{\beta i} = 0$, which is expected, also that the order of magnitudes are finite at $T_g$. Next we plot the inverse relaxation times defined equations(9).

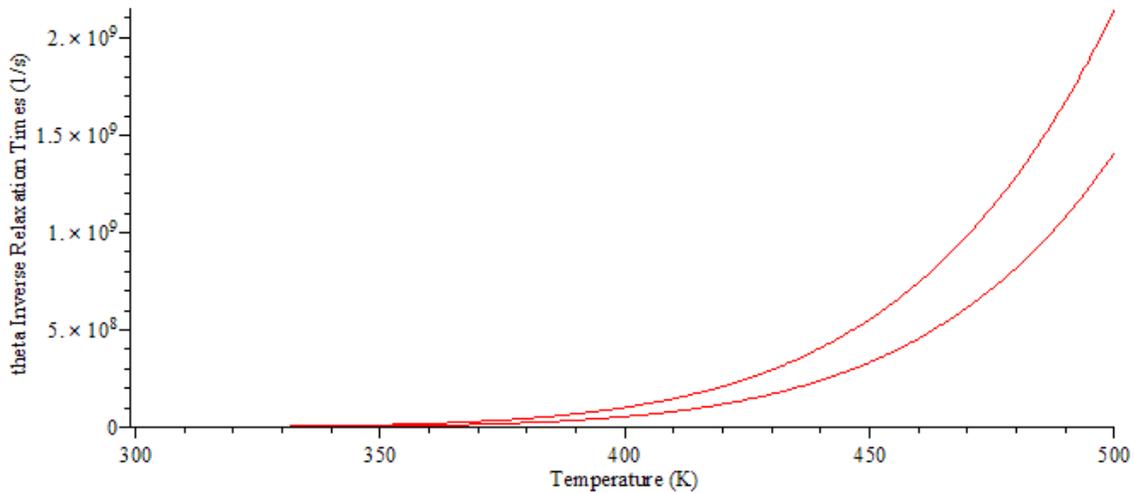

**Figure 5. Inverse relaxation times $\gamma_{\theta 11}, \gamma_{\theta 12}$. Note the strong overlap of these quantities as $T \to 0$.**

Notice that these also converge: $\lim_{T\to 0}\gamma_{\theta ij} = 0$, which is expected, also that the order of magnitudes are finite at $T_g$. Both sets of inverse relaxation times show approximately the same behaviors.

One subtlety of this calculation is that the energy associated with each fluctuation freezes out at a slightly different time. Since the ground state energy of each bonding configuration is unique, and we can put the time dependence into equation (1) as,

$$\delta F_{\beta_i}(t) = \left\{ \left( \sum_i^j \frac{\partial E_0(\beta_i)}{\partial \beta_i} - \left(\Phi_{\beta i}\right)^2 \cdot d_i \right) \delta O^2{}_{0\beta i} e^{-2\gamma_\beta t} - \kappa d_i^2 q_{0\beta i} \delta O^2{}_{0\beta i} e^{-2\gamma_\beta t} + \frac{\kappa d_i^2}{2} \delta O_{0\beta i}{}^4 e^{-4\gamma_\beta t} \right\}$$
.(10)

Here we take the real part of both sides to eliminate the imaginary time dependence of the emitted bath quanta $q_{0\beta i}$. We have shown this ground state energy is unique, we assume that there is some characteristic time $\Delta t_\beta$ such that,

$$\left. \frac{\partial \left(\delta F_{\beta_i}(t)\right)}{\partial t} \right|_{\Delta t_\beta} \approx 0 \tag{11}$$

This time can be solved for as,

$$\Delta t_{\beta i} = \frac{1}{2\gamma_{\beta i}} \ln \left\{ \frac{\left(\kappa d_i^2 \delta \beta_{0i}\right)}{\left[ \left( \sum_i^j \frac{\partial E_0(\beta_i)}{\partial \beta_i} - \left(\sqrt{|\vec{f}_i|} + i\sqrt{\hat{f}_i \Box \sum_{j\neq i} \vec{f}_j}\right)^2 \cdot d_i \right) - \kappa d_i^2 q_{0\beta i} \right]} \right\}. \tag{12}$$

This is the time for the energy of the bond fluctuations to stabilize into the ground state, and as such is conjugate to the energy $\delta F_{\beta_i}(t)$.

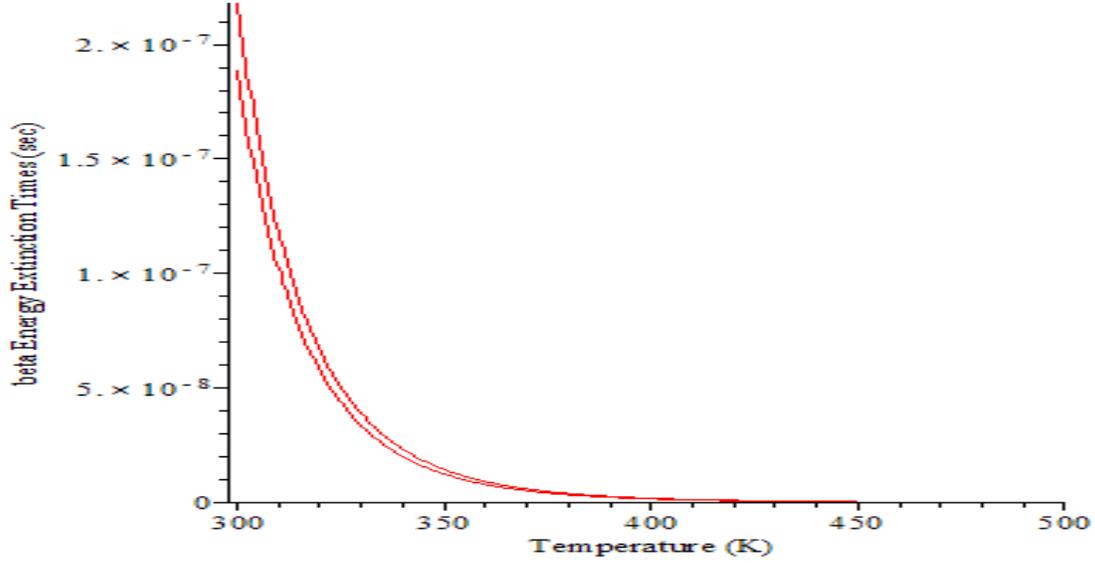

**Figure 6.** Energy Extinction Time for $\delta F_{\beta_i}(t)$.

Note this converges as: $\underset{T\to\infty}{Lim}\Delta t_{\beta i}=0$ as one would expect. Also note that the fluctuation $\delta F_{\beta_i}(t)$ dies out above $T_g$. With the exception of the inverse relaxation times, all of the quantities in equation (12) are determined by equations (3). Similarly, the time dependent equation for the energy fluctuations in $\delta \theta_i$ can be written as,

$$\delta F_{\theta_{ij}}(t) = \left\{ \frac{\partial E(\theta_{ij})}{\partial \theta_{ij}} \delta\theta_{0ij} e^{-\gamma_{ij\theta}t} + \sum_i \frac{\partial\left[V_{pp\sigma,\pi i}\right]}{\partial \theta_{ij}}\delta\theta_{0ij} e^{-\gamma_{ij\theta}t} - \kappa d^2 q_{0\theta}\delta\theta_{0ij} e^{-\gamma_{ij\theta}t} + \frac{\kappa d^2}{2}\delta\theta_{0ij}^{\ 2} e^{-2\gamma_{ij\theta}t} \right\}$$

.(13)

Here as in equation (10) we have taken the real part to eliminate the imaginary time dependence of the quantity $q_{0\theta}$. We again assume that this fluctuation in the energy is also stationary at some time $\Delta t_{\theta_{ij}}$ then we obtain,

$$\left.\frac{\partial\left(\delta F_{\theta_{ij}}(t)\right)}{\partial t}\right|_{\Delta t_{\theta ij}} \approx 0 \qquad (14)$$

$$\Delta t_{\theta ij} = \frac{1}{2\gamma_{\theta ij}} \ln\left\{ \frac{\left(\kappa d_i^2 \delta\theta_{ij}\right)}{\left(\left(\frac{\partial E_0(\theta_{ij})}{\partial \theta_{ij}} + \sum_{ij}\frac{\partial\left[V_{pp\sigma,\pi i}\right]}{\partial \theta_{ij}}\right) - \left(\kappa d_i^2 q_{\theta ij}\right)\right)} \right\}. \qquad (15)$$

This is plotted below:

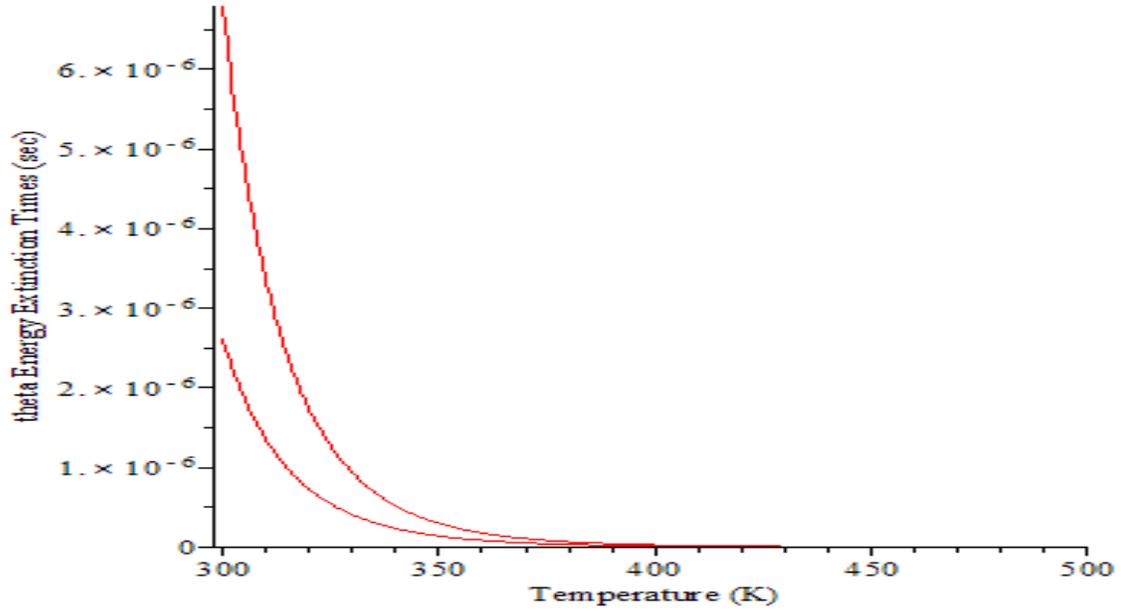

**Figure 7. Energy Extinction Time for** $\delta F_{\theta ij}(t)$

Note this converges as: $\underset{T\to\infty}{Lim}\Delta t_{\theta ij}=0$ as one would expect. Also note that the fluctuation $\delta F_{\theta ij}(t)$ dies out above $T_g$. With the exception of the inverse relaxation times, all of the quantities in equation (15) are determined by equations (3). Note that the extinction times shown in figure (6) are an order of magnitude larger than those in figure (7) around $T_g$. Of crucial importance is the fact that the time intervals $\gamma_{\beta i}^{-1}, \gamma_{\theta ij}^{-1}$ required for the fluctuations $\delta\beta_{0i}, \delta\theta_{0ij}$ to freeze out are <u>different</u> than the corresponding energy extinction times $\Delta t_{\beta i}, \Delta t_{\theta ij}$. Based on the time dependence inferred from Figures 4 & 5 we identify $\gamma_{\beta i}^{-1}, \gamma_{\theta ij}^{-1}$ with fast relaxation processes [8]. The self organized criticality approach can be understood by considering Figure 8 below:

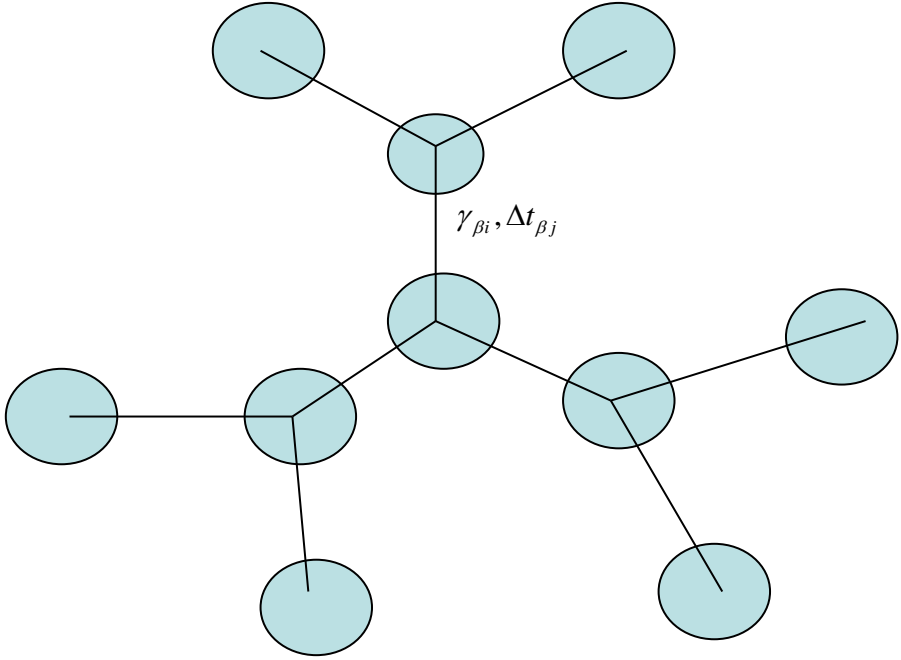

**Figure 8. the smallest complete set of bonding LPS pyramidal units. Blue dots correspond to As atoms.**

Here each line in Figure 8 corresponds to a different As-As bond. This structure is defined in terms of bond fluctuations which have the time dependence described in eqns. (8) & (9). As such it shares some properties with an STZ as defined in ref.[ 1 ], albeit with the frustration forces and noise fluctuations from the bath competing to "freeze in" irreversible molecular arrangements.

Here each line in Figure 8 corresponds to a different As-As bond. Following reasoning similar to that of Per Bak [21, 22], consider a collection of LPS as shown above, with all possible different combinations of the products $\gamma_{\beta i} \Delta t_{\beta j}$. Using equations (8,12) we define the sums,

$$\bar{\gamma}_\beta = \frac{\sum_i^3 \gamma_{\beta i}}{3}, \bar{\Delta t}_\beta = \frac{\sum_i^3 \Delta t_{\beta i}}{3}. \quad (17.a,b)$$

The product $\left[\bar{\gamma}_\beta \bar{\Delta t}_\beta\right]$ would correspond to the normalized "length" of a collection of LPS. We can define a dimensionless "length" of a single LPS as,

$$\left(\gamma_{\beta i} \Delta t_{\beta j}\right) = \sum_{i,j} \gamma_{\beta i} \Delta t_{\beta j} \delta_{ij}. \quad (18)$$

Assuming the structure shown above is self organizing, we posit a relationship between the quantities defined in equations (17, 18) as,

$$\left[\bar{\gamma}_\beta \bar{\Delta} t_\beta \right] = \left(\gamma_{\beta i} \Box \Delta t_{\beta j}\right)^{\xi_\beta(T)} . \tag{19}$$

To see that this quantity is an example of self organized criticality we define,

$$\eta_{\beta i} = \left\{ \frac{\left(\kappa d_i^2 \delta \beta_{0i}\right)}{\left[\left(\sum_i^j \frac{\partial E_0(\beta_i)}{\partial \beta_i} - \left(\sqrt{|\vec{f}_i|} + i\sqrt{\hat{f}_i \Box \sum_{j \neq i} \vec{f}_j}\right)^2 \cdot d_i\right) - \kappa d_i^2 q_{0\beta i}\right]} \right\}. \tag{20}$$

This quantity is a scalar with no explicit temperature dependence. A sum of these, $\sum_i \ln[\eta_{\beta i}]$, represents the $\delta \beta_i$ part of a particular LPS in figure 8 above.

Equation (20) can now be written as,

$$\sum_j \frac{\gamma_{\beta j}(T)}{\gamma_{\beta i}(T)} \left(\sum_i \ln[\eta_{\beta i}]\right) = 9 \left(\sum_i \ln[\eta_{\beta i}]\right)^{\xi_\beta(T)}. \tag{21}$$

The left hand side is a temperature dependent distribution of these sums, and as such is representative of self organized criticality [9]. $\xi_\beta(T)$ can solved for in terms of known quantities,

$$\xi_\beta(T) = \frac{\ln\left[\left(\sum_i \gamma_{\beta i}\right)\left(\sum_j \Delta t_{\beta j}\right)\right] - \ln(9.0)}{\ln\left[\sum_i^3 \left(\gamma_{\beta i} \Delta t_{\beta j} \delta_{ij}\right)\right]}. \tag{22}$$

This will be evaluated in the conclusion. We do a similar construction for the fluctuations $\delta \theta_{0ij}$,

$$\bar{\gamma}_\theta = \frac{\sum_{i,j}^{3} \gamma_{\theta ij}}{3}, \bar{\Delta t}_{\theta ij} = \frac{\sum_{i}^{3} \Delta t_{\theta ij}}{3}. \tag{23}$$

We similarly define a type of dimensionless inner product as,

$$\left(\gamma_{\theta ij} \Box \Delta t_{\theta ik}\right) = \sum_{i,k} \gamma_{\theta ij} \Delta t_{\theta ik} \delta_{ik}. \tag{24}$$

Again defining a scalar quantity representing a particular LPS,

$$\eta_{\theta ij} = \left\{ \frac{\left(\kappa d_i^2 \delta\theta_{ij}\right)}{\left[\left(\frac{\partial E_0(\theta_{ij})}{\partial \theta_{ij}} + \sum_{ij} \frac{\partial \left[V_{pp\sigma,\pi i}\right]}{\partial \theta_{ij}}\right) - \left(\kappa d_i^2 q_{\theta ij}\right)\right]} \right\}. \tag{25}$$

This quantity is also a scalar with no explicit temperature dependence. A sum of these, $\sum_i \ln\left[\eta_{\theta ij}\right]$, represents the $\delta\theta_{ij}$ part of a particular LPS in figure 8 above. Again we obtain,

$$\sum_j \frac{\gamma_{\theta ij}}{\gamma_{\theta ik}} \left(\sum_i \ln\left[\eta_{\theta ik}\right]\right) = 9 \left(\sum_i \ln\left[\eta_{\theta ik}\right]\right)^{\xi_\theta(T)}. \tag{26}$$

And,

$$\xi_\theta(T) = \frac{\ln\left[\left(\sum_{i,j}\gamma_{\theta ij}\right)\left(\sum_{j,k}\Delta t_{\theta jk}\right)\right] - \ln(9.0)}{\ln\left[\sum_i^3 \left(\gamma_{\theta ij}\Delta t_{\theta ij}\delta_{ij}\right)\right]}. \tag{27}$$

Here we note that the bath modes $q_{\beta i}, q_{\theta ij}$ which in part determine equations (20) and (25) also in part produce the observed NQR distribution [10]. In the conclusion we show that $0.5\left\{\xi_\theta\left(T_g\right)^{-1} \approx \xi_\beta\left(T_g\right)^{-1}\right\} \approx 0.6$.

## III. Conclusions

Using a self organization approach We obtained the two power law terms, $\xi_\theta(T), \xi_\beta(T)$. These are plotted below.

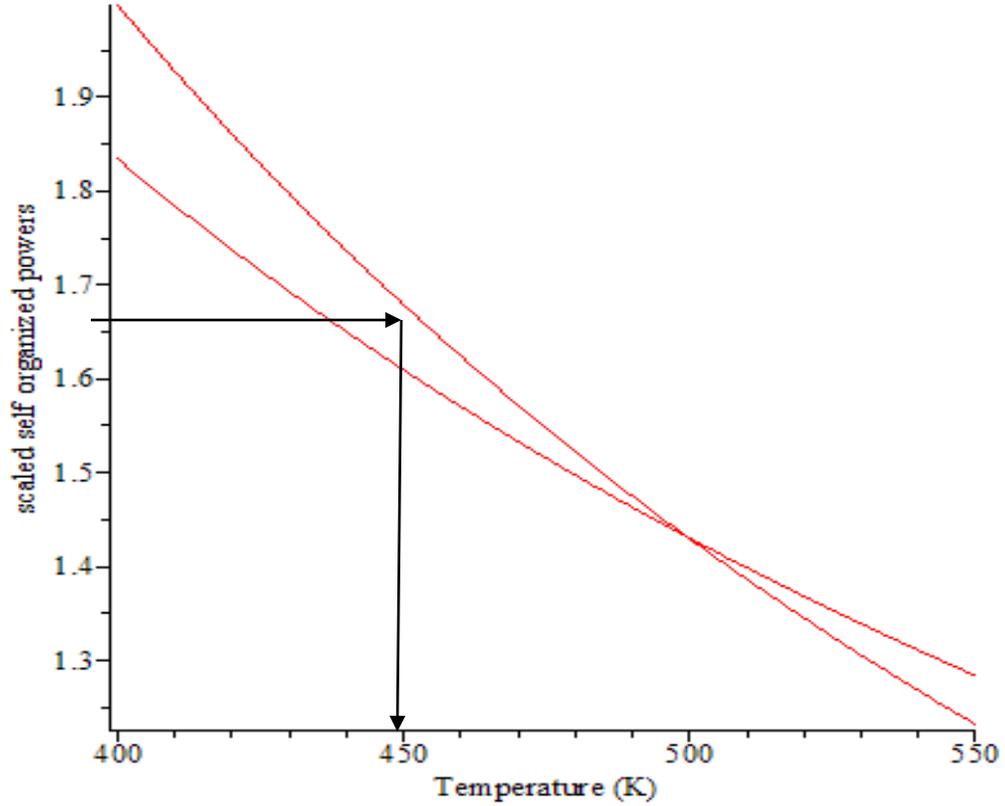

**Figure 9. plots of $\xi_\theta(T), \xi_\beta(T)$ as functions of temperature (Kelvin).**

Each of these two quantities emerge as a result of the relationship between the fluctuation relaxation time and the energy extinction time of a given PLS. Note first that the two quantities cross at approximately 500 K, which is just above $T_g$, and that the magnitudes are small. The magnitudes of $\xi_\theta(T), \xi_\beta(T)$ can be understand in terms of self organized criticality, while the convergence,

$$\underset{T \to T_g}{Lim}\left[\xi_\theta(T) \approx \xi_\beta(T)\right]. \tag{28}$$

has its origin in gauge freedom. Below we plot a scaled version of the inverse of these, $\kappa\xi_\beta(T)^{-1}, \kappa\xi_\theta(T)^{-1}$,

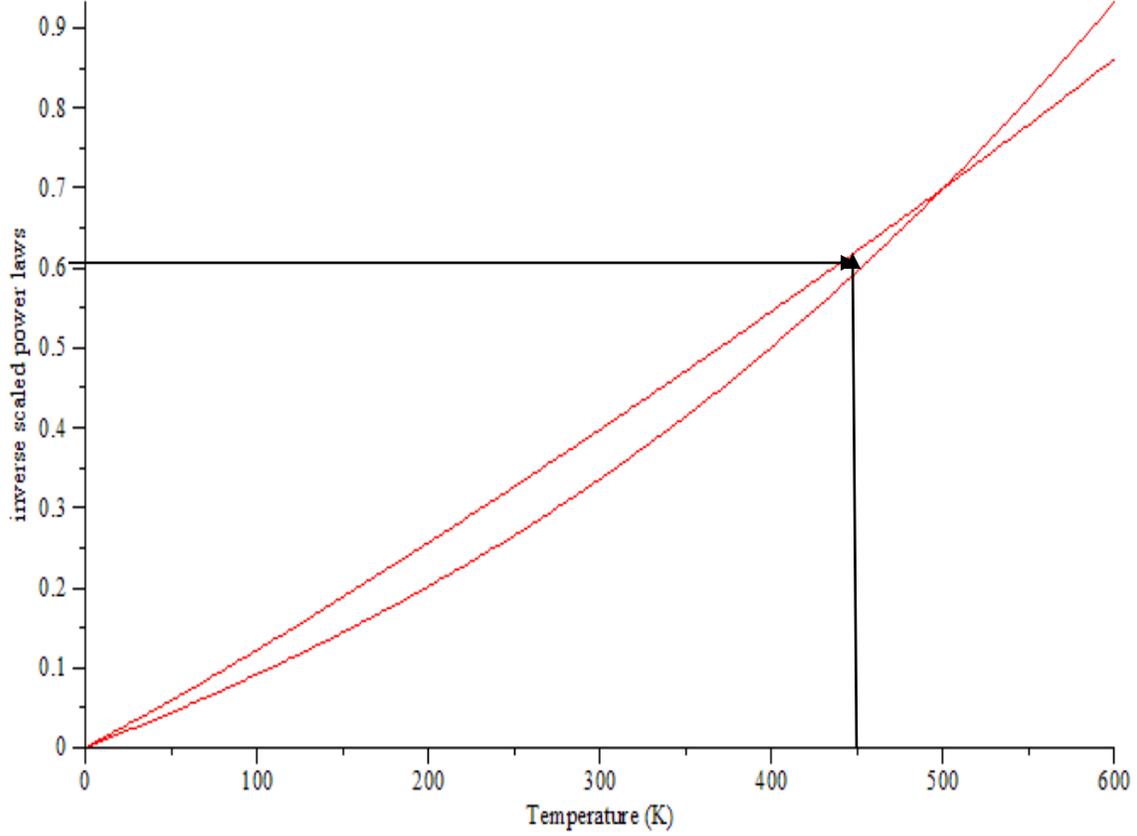

**Figure 10. plots of $\kappa\xi_\beta(T)^{-1}, \kappa\xi_\theta(T)^{-1}, \kappa = 0.5$ as functions of temperature.**

Note that these converge as the system approaches the glass temperature. Here $\kappa = 0.5$.

$$\lim_{T \to T_g}\left[\xi_\theta(T)^{-1} \approx \xi_\beta(T)^{-1}\right] \approx 0.6 = \lambda_{glass}. \tag{30}$$

We posit that self organized criticality is responsible for these two quantities to both converge to 0.6 at the glass temperature. In the stretched exponential function $\phi(t) = \exp\left[(-t/\tau)^\lambda\right]$ $\lambda$ has as an argument the product $(t/\tau)$, which is dimensionless. Similarly, $\xi_\theta(T), \xi_\beta(T)$ have as their arguments the products $(\gamma_{\beta i}\Box\Delta t_{\beta j}), (\gamma_{\theta ij}\Box\Delta t_{\theta ik})$, which are also dimensionless in terms of the same variable time. Since $\lambda$ is typically not derived as a function of temperature, we note the fractal nature of $\lambda$ as derived here can be seen as,

$$\xi_{\beta,\theta}\left(T_g\right) \approx \frac{1}{\lambda} = 1 - D, \rightarrow D = \frac{\lambda - 1}{\lambda} = -0.66 . \tag{31}$$

Mandelbrot [30] specifies that "negative fractal dimensions describe the fluctuations one may see in a finite size sample". Since this is a model of fluctuations equation (31) is consistent with $\xi_\theta\left(T_g\right) \approx \xi_\beta\left(T_g\right) \approx \lambda^{-1}$ being emergent quantities which are a result of cooperative self organization. We note that the structure shown in figure 8 is the smallest "self organizing" configuration one can use to calculate $\lambda$, and as such may place a lower limit on the size of relevant structures defined on a larger scale, such as in STZ and RFOT theory.